\documentclass[letterpaper,seceq]{jpsj2}
\newlength{\figwidth}
\newlength{\currentfigwidth}
\title{Deconfined criticality critically defined}

\author{T. \textsc{Senthil}$^{1}$, Leon \textsc{Balents}$^{2}$, Subir
  \textsc{Sachdev}$^{3}$,\\ Ashvin
\textsc{Vishwanath}$^{1}$, and Matthew P. A.
\textsc{Fisher}$^{4}$}

\inst{$^{1}$Department of Physics, Massachusetts Institute of
  Technology, Cambridge MA 02139 \\
$^{2}$Department of Physics, University of California, Santa Barbara CA
93106-4030 \\
$^{3}$Department of Physics, Yale University, P.O. Box 208120, New
Haven CT 06520-8120\\
$^{4}$Kavli Institute for Theoretical Physics, University of
California, Santa Barbara, CA 93106-4030}

\abst{We describe characteristic physical properties of the recently
  introduced class of deconfined quantum critical points. Using some
  simple models, we highlight observables which clearly distinguish such
  critical points from those described by the conventional
  Landau-Ginzburg-Wilson framework: such a distinction can be made quite
  precisely even though both classes of critical points are strongly
  coupled, and neither has sharp quasiparticle excitations.  We also
  contrast our classification from proposals by Bernevig {\em et al.}
  and Yoshioka {\em et al.}.  \\~\\
  {\tt Proceedings of the International Conference on Statistical
Physics of Quantum Systems -- novel orders and dynamics, July
17-20, 2004, Sendai, Japan}}

\kword{antiferromagnet, valence bond solid, spin liquid, gauge
theory, topological order}

\begin{document}
\maketitle

\section{Introduction}

A major focus of the study of correlated electron systems has been
the analysis of continuous quantum phase transitions in two
spatial dimensions. These are described by quantum field theories
in 2+1 spacetime dimensions, and the most interesting are
strongly-coupled: this means that there is no formulation of the
critical theory in terms of free boson or fermion fields. (Indeed
no such free field description is known for a wide variety of
interesting critical points in $2 +1$ dimensions, and most probably
simply does not exist). Rather, the
gapless critical excitations interact with each other with a
coupling of order unity. In contrast, in 1+1 dimensions it is
often the case that there is a preferred formulation of the
critical theory in terms of free fields. In the absence of such a
free field formulation, there is no direct and simple physical
interpretation of the excitation spectrum of 2+1 dimensional
critical points, and of the interactions between the excitations.

Nevertheless, a great deal of understanding of some 2+1
dimensional critical points has been achieved by what may be
called the Landau-Ginzburg-Wilson (LGW) framework \cite{ma}. In
the LGW approach one identifies the order parameter (denoted
$\varphi$) and elementary excitations (usually, fluctuations of
$\varphi$) characterizing the phases adjacent to the critical
point; then, as we review below, a sophisticated renormalized
perturbation theory of the interactions between $\varphi$
excitations can be developed to also describe the spectrum of the
quantum critical point.

In recent work \cite{shortpap}, we have argued that there are
certain 2+1 dimensional critical points which do not fall into the
above LGW framework. For these critical points, the best starting
point for a description of the critical theory is not the order
parameter, but an emergent set of fractionalized excitations which
are special only to the critical point, and not present in either
phase adjacent to the critical point. Moreover, there is an
additional {\em topological structure\/} present at the critical
point, connected with a topological conservation law. The extra
conserved quantity is conveniently interpreted as the total flux
of a gauge field that emerges at the critical point. This total
flux is conserved only asymptotically at low energies at the
quantum critical point. This conservation law provides a sharp
distinction between these {\em deconfined\/} critical points and
the conventional LGW critical points (for which there is no such
conservation law). We will review an example of such a deconfined
critical point below.

We strongly emphasize that the phenomena displayed near the
quantum critical points we have dubbed deconfined are strikingly
different from those near more conventional (even if strongly
interacting) ones. Thus the nomenclature `deconfined' has physical
content, and is not just a matter of imposing terminology on some
familiar and well-understood critical phenomena. Indeed, not all
strongly interacting critical points are to be regarded as
deconfined.

In earlier work, Laughlin \cite{rbl} had proposed
fractionalization at 2+1 dimensional critical points on
phenomenological grounds. However, the specific implementation of
his scenario by Bernevig, Giuliano, and Laughlin \cite{bgl}
disagrees with the classification we have proposed above. Indeed
the example considered - the usual O($n$) critical point in $2+1$
dimensions, the field theory in Eq.~(\ref{sp}) below - is
universally accepted as the canonical example of LGW criticality,
including by Bernevig {\em et al.} \cite{bgl}. Despite this,
Bernevig {\em et al.} \cite{bgl} predicted ``a new spectroscopic
effect that should occur very generally at quantum phase
transitions described by O($n$) $\sigma$-models''---the existence
of ``excitations similar to meson resonances'' of fractionalized
spin $S=1/2$ ``elementary excitations with integrity analogous to
spinons''.  As discussed in Section~\ref{lgw}, for the usual
O($n$) critical point as described by Eq.~(\ref{sp}), the spectrum
\cite{csy} near the phase transition can be fully understood in
terms of the renormalized perturbation theory of $S=1$ $\varphi$
excitations, and it does not display any signatures of
fractionalized $S=1/2$ spinon-like excitations at any length
scale. In particular, the resonances predicted by Bernevig {\em et
al.} simply do not exist in this model. Moreover, if such
resonances are found for a particular model which exhibits a
quantum critical point in this universality class, they must be
ascribed to non-universal (usually) lattice scale phenomena
unrelated to the universal critical singularities associated with
the phase transition.

An example of our deconfined critical points is reviewed in
Section~\ref{deconfine}: it is described by the critical quantum
field theory in Eq. (\ref{zc}) of $S=1/2$ complex spinors $z_a$
and a non-compact U(1) gauge field $A_\mu$. The presence of the
gauge field, and the associated `photon'-like excitations, is a
consequence of the topological conservation law at the critical
point. A key point is that this field theory applies at all length
scales (much larger than the lattice spacing) only at the quantum
critical point. Away from the critical point, there is a very
large length scale (much larger than the usual order parameter
correlation length) beyond which `dangerously irrelevant'
corrections eventually dominate, and lead to the loss of the
topological conservation law. The fractionalized degrees of
freedom that rear their head at the critical point also undergo
confinement by this scale. The presence of this second very large
length scale is an important secondary characteristic of our
deconfined critical points.

On a different note, we also strongly emphasize that the
occurrence of deconfined quantum critical points does not in the
least imply that fractionalized excitations are {\em necessarily}
associated with quantum criticality in $2+1$ dimensional strongly
correlated systems (other than in the fractional quantum Hall
effect). Indeed, a large body of solid theoretical work over the
last several years has established the stability of quantum {\em
phases} of matter in two or higher dimensions where the
excitations have fractional quantum numbers. These higher
dimensional fractionalized phases, and the deconfined quantum
critical points, have in common the presence of extra topological
structure absent in the microscopic models in which they arise.
This structure is associated with the presence of new topological
conservation laws that are useful characterizations of the
deconfinement. The corresponding conserved quantities can
fruitfully be viewed as the fluxes of appropriate deconfined gauge
fields.

We will present examples of a LGW and a deconfined critical point
in the following two sections, and contrast their properties. A
simple example of a LGW critical point, presented in
Section~\ref{lgw}, is provided by lattice models of O($n$) quantum
rotors. These display second order quantum phase transitions
between ordered phases where the O($n$) symmetry is broken, and
simple disordered phases which preserve all the symmetries of the
microscopic Hamiltonian. The phase transition is in the
universality class of the usual O($n$) model in 2+1 dimensions
(and is described by the continuum O($n$) non-linear sigma model
field theory in Eqs.~(\ref{defQ}) and (\ref{sp})). A more
sophisticated microscopic situation, presented in
Section~\ref{sec:dimer}, which displays a quantum transition in
the same universality class is a model of spin-$1/2$ quantum
Heisenberg spins on a `dimerized' lattice. In all these cases the
transition is well-described within the standard LGW paradigm.
Indeed this paradigm was evolved in the specific context of these
universality classes - albeit viewed as thermal phase transitions
in three dimensions. In contrast to this situation, our deconfined
critical points\cite{shortpap}, reviewed in
Section~\ref{deconfine}, display strikingly different phenomena.
We will review one example - namely the transition between
N\'{e}el and valence bond solid ordered phases of spin-$1/2$
square lattice quantum antiferromagnets - that aptly illustrates
the breakdown of the LGW paradigm.

The following discussion in this section presents characteristics
of the phase transition which apply in {\em both} the LGW and
deconfined cases.

One of the phases beside both phase transitions is characterized
by the non-zero expectation value of the N\'{e}el order parameter
$\varphi_{\alpha}$ ($\alpha=x,y,z$), representing the component of
the staggered spin polarization at an ordering wavevector ${\bf
K}$. We tune the system across the quantum phase transition by
varying a coupling $g$. For $g<g_c$, the ground state has N\'{e}el
order and hence
\begin{equation}
\langle \varphi_{\alpha} \rangle \neq 0~~~~~~;~~~~~\mbox{$g<g_c$,
N\'{e}el state.} \label{eq:neel}
\end{equation}
The quantum critical point of interest is at $g=g_c$, and for
$g>g_c$ SU(2) spin rotation invariance is restored and we have the
paramagnetic ground state:
\begin{equation}
\langle \varphi_{\alpha} \rangle = 0~~~~~~;~~~~~\mbox{$g>g_c$,
paramagnetic state.} \label{eq:para}
\end{equation}
In both the cases we consider below, the paramagnetic state has a
sharp $S=1$ excitation which can be interpreted as the
oscillations of the order parameter $\varphi_{\alpha}$ about its
zero mean value in Eq. (\ref{eq:para}). This implies that the
susceptibility of $\varphi_{\alpha}$ autocorrelations,
$\chi_\varphi$, has the following form at small frequencies
($\omega$) and wavevectors ($k$):
\begin{equation}
\mbox{Im} \left[ \chi_\varphi (k, \omega) \right] \sim
\frac{Z}{\Delta} \delta(\omega - \Delta - \mathcal{O} (k^2)) +
\ldots~~~;~~~\mbox{$g>g_c$}. \label{triplon}
\end{equation}
Here $\Delta$ is the spin gap above which the $S=1$ excitation has
a quadratic dispersion,and $Z$ is the quasiparticle residue. The
low energy spinful excitations of the paramagnet therefore have a
simple quasiparticle interpretation.

Now let us consider the spectrum at the critical point $g=g_c$.
Both critical points described below have an effective
`relativistic' invariance and the critical susceptibility has the
form
\begin{equation}
\chi_\varphi (k, \omega) \sim \frac{1}{(c^2 k^2 -
\omega^2)^{1-\eta/2}} ~~~;~~~\mbox{$g=g_c$},\label{chicrit}
\end{equation}
where the exponent $\eta$ is anomalous dimension of the
$\varphi_{\alpha}$ field, and $c$ is a velocity. Note that because
$\eta \neq 0$, the imaginary part of Eq. (\ref{chicrit}) does not
have quasiparticle delta-function, but only a continuum
contribution for $\omega > c k$. So, clearly, there are no $S=1$
quasiparticle excitations at the quantum critical point. This
absence of quasiparticles is a general property of
strongly-coupled quantum critical points.

In Section~\ref{lgw}, we will argue that in many cases the
continuum spectral density in Eq. (\ref{chicrit}) can be
understood in LGW theory by an evolution from the quasiparticle
spectral density in Eq. (\ref{triplon}) (contrary to claims by
Bernevig {\em et al.}\cite{bgl}). We can use perturbative
renormalization group techniques to analyze the consequences of
interactions between the $S=1$ quasiparticle excitations. As we
approach the critical point with $g \searrow g_c$, these
interactions lead \cite{csy} to a decrease in $Z$ until it
vanishes at $g=g_c$, and the spectral density takes the
non-quasiparticle form in Eq. (\ref{chicrit}). The magnitude of
the anomalous dimension $\eta$ is usually small in such an
approach, and this underlies the success of the perturbative
approach.

Section~\ref{deconfine} will turn to the novel deconfined critical
points. The forms in Eqs. (\ref{eq:neel}), (\ref{eq:para}),
(\ref{triplon}), and (\ref{chicrit}) continue to apply in this
case too. So the critical spectrum in Eq. (\ref{chicrit}) is
present in both the LGW and deconfined cases, and is {\em not\/}
an indicator of fractionalization by itself. However in the
deconfined case, the underlying theory of the excitations leading
to the response function in Eq. (\ref{chicrit}) is different, and
expressed in terms of fractionalized modes. A sharp observable
distinction between the two cases lies in the emergence of a new
conservation law at the critical point. The topological flux is
defined (in the continuum limit) by
\begin{equation}
\mathcal{Q} = \frac{1}{4 \pi} \int dx dy
\epsilon_{\alpha\beta\gamma} \varphi_{\alpha}
\partial_x \varphi_\beta
\partial_y \varphi_\gamma ,\label{defQ}
\end{equation}
(here, we have rescaled $\varphi_{\alpha}$ to be a unit
vector---details in the sections below) and measures the skyrmion
number \cite{rajaraman} of the spin configuration. $\mathcal{Q}$
is conserved only at the $g=g_c$ deconfined critical point, and
not at other values of $g$. It is also not conserved at {\em any} value
of $g$ for the LGW case discussed in Section~\ref{lgw}. The
conservation of $\mathcal{Q}$ signifies the emergence of
an extra global $U(1)$ symmetry at the deconfined critical point, and the
presence of an additional set of gapless gauge excitations. A
secondary consequence is that it naturally leads to larger
values of the exponent $\eta$ in Eq. (\ref{chicrit}) for
deconfined critical points.

\section{LGW criticality}
\label{lgw}

Consider a lattice model of O($n$) quantum rotors. The $n = 2$
case is well-studied in the context of superconductor-insulator
transitions in Josephson junction arrays. The $n = 3$ case
describes quantum phase transitions in certain classes of quantum
antiferromagnets. This class of models has been regarded as
prototypical of quantum phase transitions (in much the same way as
the classical O($n$) models are prototypes of thermal phase
transitions). Here we will briefly review their properties. This
will set the stage to appreciate the novel and unusual phenomena
near the deconfined quantum critical points.

For concreteness, we specialize to the case $n = 3$.
\begin{equation}
H = g\sum_r \frac{\vec L_r^2}{2} - \frac{1}{g}\sum_{\langle rr'
\rangle}\hat n_r \cdot \hat n_{r'}
\end{equation}
Here $\hat n_r$ is a three component unit vector on the sites $r$
of a square lattice. The vector operator $\vec L_r$ is the
corresponding angular momentum which generates rotations of $\hat
n_r$. For small $g$ the second `potential' term in the Hamiltonian
dominates and the $\hat n$ vector orders:
\begin{equation}
\langle \hat n_r \rangle \neq 0
\end{equation}
The O(3) symmetry of the Hamiltonian $H$ is then spontaneously
broken down to O(2). The low energy excitations are simply two
linearly dispersing `spin waves' as required by Goldstone's
theorem. For large $g$, on the other hand, the first term
dominates. In this case the ground state is a paramagnet which
preserves all the symmetries of the Hamiltonian. (A simple
caricature of the ground state is obtained by putting each rotor
in the $\vec L_r^2 = 0$ state.) The low energy excitations about
this state are now gapped and simply correspond to a massive
triplet of spin-$1$ bosons.

Upon increasing $g$ there is a quantum phase transition between
these two phases. The crucial conceptual idea behind the
description of this transition is that all the universal critical
singularities are due to the long wavelength long time
fluctuations of the order parameter field $\hat n$. This idea
(which can ultimately be traced to Landau) underlies all of our
understanding of phase transitions. A suitable continuum theory
that describes these fluctuations is indeed the $O(3)$ non-linear
sigma model field theory in $2 +1$ dimensions:
\begin{equation}
S_{\rm nlsm} = \int d^2x d \tau \frac{1}{2g}
\left[\left(\partial_x \hat n \right)^2 + \left(\partial_y \hat n
\right)^2 + \frac{1}{c^2}\left(\partial_{\tau}\hat n\right)^2
\right] \label{nlsm}
\end{equation}
Equivalently, we may soften the unit vector constraint on $\hat n$
by letting $\hat n \sim \vec \varphi$ and study a `soft-spin'
version of the same theory which has the same universal critical
properties:
\begin{eqnarray}
&& \mathcal{S}_{\varphi} = \int d^2 r d \tau \left[ \frac{1}{2}
\Bigl\{ \left(
\partial_{\tau} \varphi_{\alpha} \right)^2 + c^2 \left(
\partial_{x} \varphi_{\alpha} \right)^2~~~~~~~~~~~~  \right.  \nonumber
\\
&~&~~~~~~~~\left. + c^2 \left(
\partial_{y} \varphi_{\alpha} \right)^2 + s \varphi_{\alpha}^2 \Bigr\} +
\frac{u}{24} \left( \varphi_{\alpha}^2 \right)^2 \right].~~~~~~~~
\label{sp}
\end{eqnarray}
Here we have rescaled $\varphi_{\alpha}$ to fix the co-efficient
of the temporal gradient term at unity, and the quantum critical
point is tuned by varying $s$; in mean-field theory $s \sim
g-g_c$.

It is a simple matter to explore the spectrum of excitations in
$g<g_c$ and $g>g_c$ phases in powers of $u$. The results are
identical in form to those obtained above for the lattice
Hamiltonian. For $g<g_c$ we obtain a doublet of spin wave modes
with linear dispersion, while for $g>g_c$
we obtain a triplet of $\varphi_{\alpha}$ oscillations about
$\varphi_\alpha=0$ with a non-zero energy gap.

The analysis of $\mathcal{S}_{\varphi}$ at $g=g_c$ requires a
somewhat more sophisticated approach. We use the fact that the
field theory $\mathcal{S}_{\varphi}$ is actually a familiar and
well-studied model in the context of classical critical phenomena.
Upon interpreting $\tau$ as a third spatial co-ordinate,
$\mathcal{S}_{\varphi}$ becomes the theory of a classical
O(3)-invariant Heisenberg ferromagnet at finite temperatures. The
Curie transition of the Heisenberg ferromagnet then maps onto the
quantum critical point between the paramagnetic and N\'{e}el
states described above. A number of important implications for the
quantum problem can now be drawn immediately.

First, the structure of the renormalization group flows that
describe the critical fixed point is well-known. The critical
fixed point has precisely one relevant perturbation (which
describes the parameter tuning the system through the transition).
The flow away from the critical fixed point ultimately ends in
stable fixed points that characterizes either of the two phases.
There is a {\em single} diverging length scale on approaching the
transition. For instance, on the paramagnetic side this may simply
be taken to be the spin correlation length. Associated with this
there is a single diverging time scale - or equivalently a
vanishing energy scale. Again, on the paramagnetic side this may
be taken to be the gap to the triplon excitations. The theory
$\mathcal{S}_{\varphi}$ has a `relativistic' invariance, and
consequently the dynamic critical exponent must be $z=1$. The
critical fixed point may be accessed in controlled expansions in
$3 - \epsilon$ dimensions or directly in $d = 2$ in an $1/n$
expansion. Excellent numerical results for a variety of universal
properties are also available.

The spin correlation length will diverge at the quantum critical
point with the exponent \cite{landau} $\nu = 0.7048(30)$. The spin
gap of the paramagnet, $\Delta$, vanishes as $\Delta \sim (g -
g_c)^{z \nu}$, and this prediction is in excellent agreement with
numerical studies of the model $H_d$ in Eq.~(\ref{ham})
below\cite{matsumoto}.

A somewhat more non-trivial consequence of the mapping to the
classical three dimensional problem is in the structure of the
spectrum at the critical point $g = g_c$. At the Curie transition
of the classical ferromagnet it is known \cite{ma} that spin
correlations decay as $\sim 1/p^{2-\eta}$, where $p$ is the
3-component momentum in the 3-dimensional classical space. We can
now analytically continue this expression from its $p_z$
dependence in the third classical dimension to the real frequency,
$\omega$, describing the quantum antiferromagnet. This immediately
yields the general result Eq. (\ref{chicrit}); the imaginary part
of the dynamic susceptibility has the form
\begin{equation}
\mbox{Im} \chi_\varphi (k, \omega) \sim \mbox{sgn}(\omega)
\sin\left(\frac{\pi\eta}{2} \right) \frac{\theta\left(|\omega| - c
|k| \right)}{\left(\omega^2 - c^2 k^2 \right)^{1-\eta/2}}
\label{chicrit2}
\end{equation}
where $\theta$ is the unit step function. As expected, there are
no quasiparticles at the critical point, and only a dissipative
critical continuum.

On moving away from the critical point - say into the paramagnetic
phase - the spectrum crosses over into that characteristic of the
paramagnetic fixed point at a scale set by the energy gap
$\Delta$. There is a delta-function pole in the dynamic
susceptibility as in Eq.~(\ref{triplon}), and additional universal
structure at higher frequencies. Indeed (at zero momentum) the
{\em only} scale for the frequency in the universal scaling limit
is $\Delta$. The presence of this single scale implies that there
can be no structure (such as resonances) in the spectral function
that have width that is parametrically smaller than their energy -
hence there is no real sharp meaning that can be given to
statements about the existence of `spinons' with integrity near
this kind of quantum phase transition.

The key point of this section is to note that the critical results
in Eqs. (\ref{chicrit}) and (\ref{chicrit2}) can be understood
entirely within the framework of the LGW theory, as presented {\em
e.g.\/} in the book by Ma\cite{ma}. We set up a renormalization
group analysis in powers of $u$, and then compute the correlators
of the critical theory by a renormalized pertubative analysis of
the renormalization group fixed point. The only additional
subtlety is the analytic continuation to real frequencies, and
this is aided by the relativistic invariance of the underlying
theory, as we have seen above.

\subsection{Coupled dimer antiferromagnet}
\label{sec:dimer}

To see how a transition in the universality class described above
may arise in quantum antiferromagnets of spin-$1/2$ Heisenberg
moments on two dimensional lattices, it is instructive to consider
the following the ``coupled dimer'' Hamiltonian\cite{gsh} (more
detail may be found in another recent review by one us
\cite{ssmott}):
\begin{equation}
H_{d} = J \sum_{\langle ij\rangle \in \mathcal{A}} {\bf S}_i \cdot
{\bf S}_j + \frac{1}{g} J \sum_{\langle ij \rangle \in
\mathcal{B}} {\bf S}_i \cdot {\bf S}_j \label{ham}
\end{equation}
where ${\bf S}_j$ are spin-1/2 operators on the sites of the
coupled-ladder lattice shown in Fig~\ref{fig1}, with the
$\mathcal{A}$ links forming decoupled dimers while the
$\mathcal{B}$ links couple the dimers as shown.
\setlength{\currentfigwidth}{\figwidth}
\addtolength{\currentfigwidth}{2in}
\begin{figure}[tb]
\centering
\includegraphics[width=\currentfigwidth]{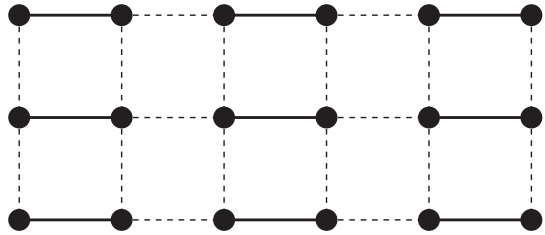}
\caption{ The coupled dimer antiferromagnet. Spins ($S=1/2$) are
placed on the sites, the $\mathcal{A}$ links are shown as full
lines, and the $\mathcal{B}$ links as dashed lines.} \label{fig1}
\end{figure}
The ground state of $H_d$ depends on the dimensionless coupling
$g$, and we will restrict our attention to $J>0$ and $g \geq 1$.
Exactly at $g=1$, $H_d$ is the familiar square lattice
antiferromagnet which describes La$_2$CuO$_4$. The $g>1$ regime
has a structure similar to models used to describe Mott insulators
like TlCuCl$_3$ \cite{cavadini,normand,oosawa}: in this case, the
value of $g$ has been tuned across the quantum critical point by
applied pressure \cite{oosawa}.

We will see that the phases and phase transitions of this model
are well represented by the O(3) quantum rotor model discussed
above. Indeed the present lattice model provides a sophisticated
example of a phase transition that fits in well with the LGW
paradigm. This important conclusion disagrees with a recent claim
by Yoshioka {\em et al.}\cite{yoshioka} of ``deconfinement'' in a
coupled dimer antiferromagnet which differs only slightly from
Eq.~(\ref{ham}). We also note that the description of coupled
dimer antiferromagnets by LGW criticality is strongly supported by
numerical studies \cite{matsumoto}.

Let us first consider the case where $g$ is close to 1. Exactly at
$g=1$, $H_d$ is known to have long-range, magnetic N\'{e}el order
in its ground state {\em i.e.} the spin-rotation symmetry is
broken and Eq. (\ref{eq:neel}) holds with
\begin{equation}
\varphi_\alpha ({\bf x}_j) = \eta_j {S}_{j\alpha} \label{neel}
\end{equation}
where $\eta_j = e^{i {\bf K} \cdot {\bf x}_j}= \pm 1$ with the
ordering wavevector ${\bf K} = (\pi,\pi)$. This long-range order
is expected to be preserved for a finite range of $g$ above 1. The
low-lying excitations above the ground state consist of slow
spatial deformations in the orientation $\langle \varphi_{\alpha}
\rangle$: these are the familiar spin waves, a standard small
fluctuation analysis yields {\em two} polarizations of spin waves
at each wavevector ${\bf k} = (k_x, k_y)$ (measured from ${\bf
K}$), and they have excitation energy
\begin{equation}
\varepsilon_k = (c_x^2 k_x^2 + c_y^2 k_y^2)^{1/2},~~~~;~~~~g<g_c
\label{spinwaves}
\end{equation}
where the spatial anisotropy of the model now requires distinct
spin-wave velocities, $c_x, c_y$, in the two spatial directions.

Let us turn now to very large $g$. Exactly at $g=\infty$, $H_d$ is
the Hamiltonian of a set of decoupled dimers, with the simple
exact ground state wavefunction shown in Fig~\ref{fig2}: the spins
in each dimer pair into valence bond singlets, leading to a
paramagnetic state which preserves spin rotation invariance and
all spatial symmetries of the Hamiltonian $H_d$.
\setlength{\currentfigwidth}{\figwidth}
\addtolength{\currentfigwidth}{2in}
\begin{figure}[tb]
\centering
\includegraphics[width=\currentfigwidth]{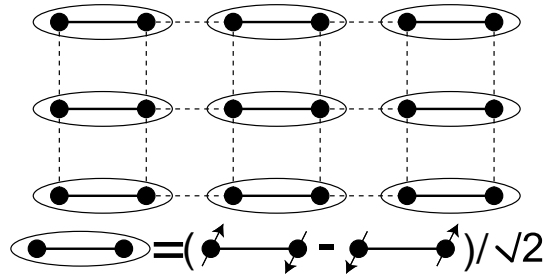}
\caption{Schematic of the quantum paramagnet ground state for $ g
\rightarrow \infty$. The ovals represent singlet valence bond
pairs. }\label{fig2}
\end{figure}
Excitations are now formed by breaking a valence bond, which leads
to a {\em three}-fold degenerate state with total spin $S=1$, as
shown in Fig~\ref{fig3}a. At $g=\infty$, this broken bond is
localized, but at small but finite $1/g$ it can hop from
site-to-site, leading to a triplet quasiparticle excitation.
\setlength{\currentfigwidth}{\figwidth}
\addtolength{\currentfigwidth}{2in}
\begin{figure}[tb]
\centering
\includegraphics[width=\currentfigwidth]{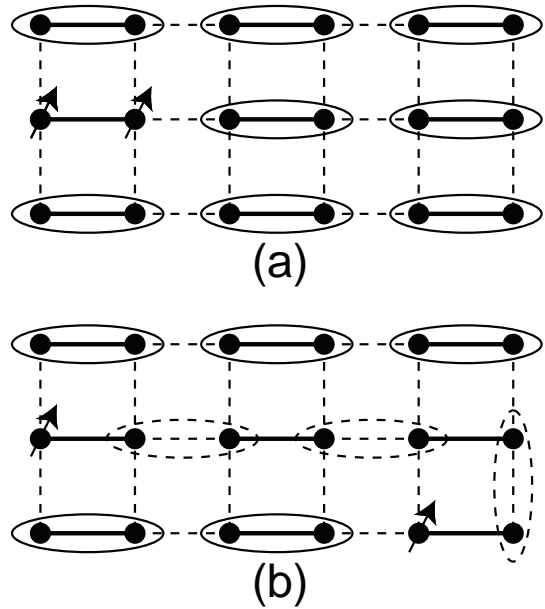}
\caption{(a) Cartoon picture of the bosonic $S=1$ excitation of
  the paramagnet. In the LGW approach, this excitation is a quantum of
  oscillation of $\varphi_\alpha$ about $\varphi_\alpha = 0$. (b)
  Fission of the $S=1$ excitation into two $S=1/2$ spinons---these are
  the $z_a$ quanta of Section~\ref{deconfine}. The spinons are connected
  by a ``string'' of valence bonds (denoted by dashed ovals) which lie
  on weaker bonds; this string costs a finite energy per unit length and
  leads to the confinement of spinons. It is important to note that this
  stringy structure, and associated excitations, is {\em
    absent\/} from the $\varphi_\alpha$ field theory in Eq.~(\ref{sp}),
  contrary to the claim by Bernevig {\em et al.} \protect\cite{bgl}. In
  this lattice spin model the ``existence" of spinons is a non-universal
  lattice scale effect, entirely unrelated to the universal long length
  scale singularities governed by the nearby quantum critical point.  By
  contrast, near the deconfined critical points of
  Section~\protect\ref{deconfine}, the stringy structure mediating
  spinon interactions is present at arbitrarily long length scales,
  being controlled by the universal critical fluctuations. But these are
  not described by the O(3) sigma model field theory of Eqs.~(\protect\ref{nlsm})
  and (\protect\ref{sp}).}
\label{fig3}
\end{figure}
Note that this quasiparticle is {\em not\/} a spin-wave (or
equivalently, a `magnon') but is more properly referred to as a
spin 1 {\em exciton} or a {\em triplon} \cite{triplon}. Indeed
this excitation is the exact analog of the gapped triplet boson in
the paramagnetic phase of the quantum rotor model. We parameterize
its energy at small wavevectors $k$ (measured from the minimum of
the spectrum in the Brillouin zone) by
\begin{equation}
\varepsilon_k = \Delta + \frac{c_x^2 k_x^2 + c_y^2 k_y^2}{2
\Delta},~~~~;~~~~g>g_c \label{epart}
\end{equation}
where $\Delta$ is the spin gap, and $c_x$, $c_y$ are velocities. A
simple perturbative calculation in $1/g$ shows that the
$\varphi_{\alpha}$ susceptibility has the form postulated in Eq.
(\ref{triplon}).

Fig~\ref{fig3} also presents a simple argument which shows that the
triplon cannot fission into two $S=1/2$ `spinons', and so the delta
function in Eq. (\ref{triplon}) is the first non-vanishing contribution
to the spectral density at low energies. For the particular quantum
phase transition discussed in the present Section, it is legitimate to
entirely neglect this meson-like structure of the triplon consisting of
a quark-antiquark pair of spinons. In particular, the `confinement'
length scale below which this structure is apparent stays finite on
approaching the transition. As such, the ``existence" of spinons and
their confining interaction in this model are non-universal lattice
scale effects, physically unrelated to the critical fluctuations that
control the universal properties of the phase transition.  In effect,
the present section contains a `chiral model' field theory of the mesons
alone, which contains no signature of the quark-like spinons, contrary
to claims by Bernevig {\em et al.} \cite{bgl}. Moreover, since the LGW
theory describes the universal critical properties of the coupled dimer
system, any signatures of such lattice-scale spinons that might obtain
in a particular model will {\sl disappear} in the vicinity of its
critical point.  By contrast, in the examples of deconfined criticality
discussed later, the confinement length scale will also diverge on
approaching the transition. It will then be necessary to include the
spinon structure of the triplon in the paramagnet in studying the
transition. We will discuss how to do this in Section~\ref{deconfine}.

The very distinct symmetry signatures of the ground states and
excitations between $g \gtrsim 1$ and $g \rightarrow \infty$ make
it clear that the two limits cannot be continuously connected. It
is known that there is an intermediate second-order phase
transition at \cite{gsh,matsumoto}  $1/g_c = 0.52337(3)$ between
these states as shown in Fig~\ref{fig4}.
\setlength{\currentfigwidth}{\figwidth}
\addtolength{\currentfigwidth}{3.5in}
\begin{figure}[tb]
\centering
\includegraphics[width=\currentfigwidth]{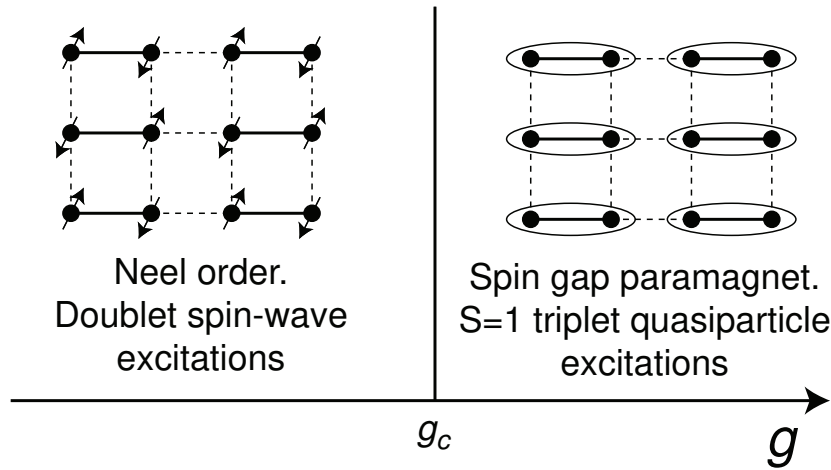}
\caption{Ground states of $H_d$ as a function of $g$. The quantum
critical point is at \protect\cite{matsumoto} $1/g_c =
0.52337(3)$. The compound TlCuCl$_3$ undergoes a similar quantum
phase transition under applied pressure \protect\cite{oosawa}
which acts to decrease the value of $g$. }\label{fig4}
\end{figure}

A quantum field theory for the quantum critical point at $g=g_c$
can be derived microscopically \cite{ssmott}, but here we guess
the answer by following the canonical LGW procedure. We focus on
the order parameter $\varphi_\alpha$, and write down the most
general effective action in powers of $\varphi_\alpha$ consistent
with lattice space group, spin rotation, and time reversal
symmetries. Such a procedure yields the familiar $\varphi^4$ field
theory in 2+1 dimensions with the action Eq.~(\ref{sp}) (a simple
spatial rescaling absorbs the anisotropic spin-wave velocities
$c_{x,y}$). Thus as far as universal critical properties are
concerned we can simply take over the discussion provided above in
the context of the rotor model.

\section{Deconfined criticality}
\label{deconfine}

In contrast to the conventional picture described in the previous
section, recent work \cite{shortpap} has discussed a number of
examples of quantum critical points which display very different
phenomena. These violate various aspects of the conventional LGW
theory. Here we will illustrate these differences in the context
of a specific example - namely a quantum phase transition between
N\'eel and valence bond solid ordered phases of spin-$1/2$
Heisenberg moments on an {\em isotropic} two dimensional lattice.

The N\'eel phase breaks spin rotation symmetry, is well-known, and
needs little further description here. The valence bond solid
(VBS) phase is a paramagnet that has some similarities with the
paramagnet in the coupled dimer Hamiltonian of
Section~\ref{sec:dimer}. However on the isotropic square lattice,
the valence bond solid state {\em spontaneously} breaks the space
group symmetry of the square lattice. A specific example of such a
VBS state is given by the columnar state of Fig.~\ref{fig7}.

Remarkably, we have argued \cite{shortpap}, it is possible to have
a direct second order phase transition between these two phases.
This is the first sign of the breakdown of the LGW paradigm. The
two phases break distinct spin rotation and lattice symmetries,
and LGW theory predicts\cite{aharony} that such states cannot
generically be separated by a continuous phase transition. The
theory of Senthil {\em et al.}\cite{shortpap} shows that there is
indeed a complete breakdown of the basic ideas of the LGW paradigm
and an essentially new description is needed of the critical
singularities. This description involves `deconfined'
fractionalized degrees of freedom in a precise sense - hence the
terminology `deconfined quantum critical points'.

Consider then spin-$1/2$ models which preserve all the symmetries
of the underlying square lattice in the class
\begin{equation}
H_s = J\sum_{\langle ij\rangle} {\bf S}_i \cdot {\bf S}_j + \ldots
. \label{hams}
\end{equation}
Here $J$ is a nearest-neighbor antiferromagnetic exchange and
takes equal values between all nearest neighbors. The ellipses
represent further short-range exchange interactions (possibly
involving multiple spin ring exchange) which preserve the full
symmetry of the square lattice. (The model $H_d$ is a member of
the class $H_s$ only at $g=1$). Here we will continue to denote by
$g$  the strength of these additional non-nearest neighbor
couplings that preserve the square lattice symmetry: an example is
shown in Fig~\ref{fig5}.
\setlength{\currentfigwidth}{\figwidth}
\addtolength{\currentfigwidth}{1.5in}
\begin{figure}[tb]
\centering
\includegraphics[width=\currentfigwidth]{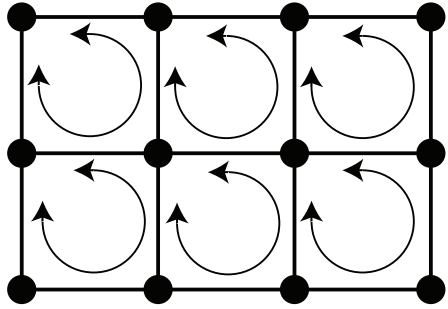}
\caption{A schematic representation of the model $H_s$ with
additional ring exchange couplings which are tuned by the value of
$g$. Unlike Fig~\protect\ref{fig1}, all couplings have the full
square lattice symmetry.} \label{fig5}
\end{figure}

A sketch of a section of the  phase diagram of $H_s$, along with
possible VBS orders in the $g>g_c$ paramagnet for $S=1/2$, is
shown in Fig~\ref{fig7}.
\setlength{\currentfigwidth}{\figwidth}
\addtolength{\currentfigwidth}{3.5in}
\begin{figure}[tb]
\centering
\includegraphics[width=\currentfigwidth]{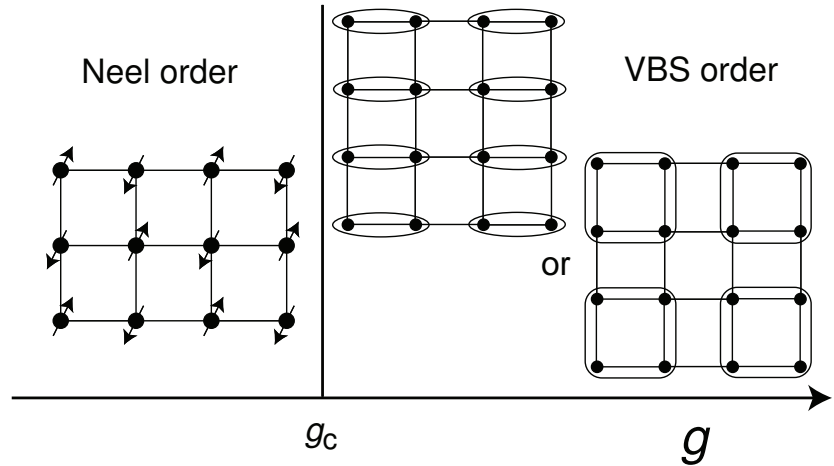}
\caption{Phase diagram of $H_s$
for $S=1/2$. Unlike the paramagnet in Fig~\protect\ref{fig4}, the
paramagnet here has spontaneous VBS order because all bonds are
equivalent in the Hamiltonian, and its ground state breaks spatial
lattice symmetries. The spinful excitations of the paramagnets in
Figs~\ref{fig4} and above are however similar, and both obey Eq.
(\protect\ref{triplon}).} \label{fig7}
\end{figure}
Note that the VBS wavefunction is actually similar to the simple
coupled dimer wavefunction in Fig~\ref{fig2}. However, because of
the absence of dimerization in the underlying Hamiltonian, this
ground state is at least four-fold degenerate. The choice among
these states leads to a particular pattern in the modulation of
the exchange energies of the bonds, and a breaking of the spatial
symmetries of the square lattice.

In the absence of the these additional couplings, it is clear that
$H_s$ has a conventional N\'{e}el ground state with the order
parameter defined in Eq. (\ref{neel}) obeying $\langle
\varphi_{\alpha} \rangle \neq 0$. Increasing the value of $g$ is
expected to enhance the quantum fluctuations about such a state,
and we attempt to describe this by setting up a coherent state
path integral over the time histories of the spins. In doing this,
it is essential to carefully account for the Berry phases of the
spins, something we circumvented in our discussion in
Section~\ref{lgw}. Explicitly, the time evolution of each spin
contributes a phase factor to the path integral given by
\begin{eqnarray}
&& \exp \Bigl( i S \times (\mbox{oriented area enclosed by
trajectory} \nonumber \\ &&~~~~~~~~~~~~~~~\mbox{of spin on the
unit sphere} ) \Bigr), \nonumber
\end{eqnarray}
where $S=1/2$ is the angular momentum of each spin; see
Fig~\ref{fig6}.
\setlength{\currentfigwidth}{\figwidth}
\addtolength{\currentfigwidth}{2in}
\begin{figure}[tb]
\centering
\includegraphics[width=\currentfigwidth]{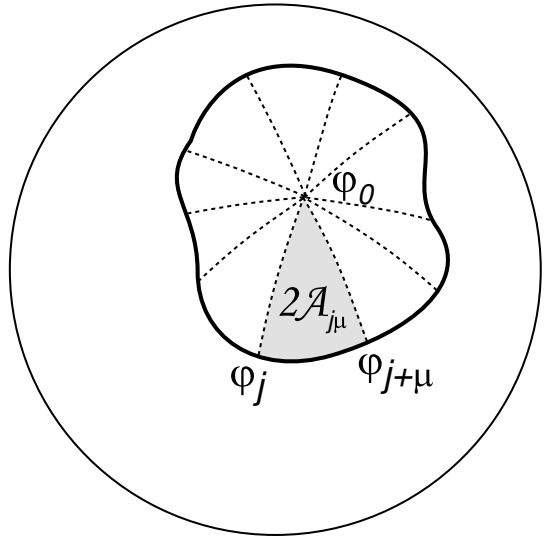}
\caption{Computation of the spin Berry phases. It is useful to
discretize spacetime into a cubic lattice of points $j$, and
define the orientation of the spin at each $j$ by a unit vector
$\varphi_{j\alpha}$ times the sublattice staggering factor
$\eta_j$ in Eq.~(\ref{neel}). Also define
$\mathcal{A}_{j\mu}=(1/2)\times$(oriented area of triangle formed
by $\varphi_{j\alpha}$, $\varphi_{j+\mu,\alpha}$ and an arbitrary
reference point $\varphi_{0,\alpha}$); here $\mu=\tau,x,y$. Then
the Berry phase contribution of all the spins is $\exp\left( i 2 S
\sum_j \eta_j \mathcal{A}_{j\tau}\right)$.} \label{fig6}
\end{figure}
Fortunately, the role of these Berry phases has been elucidated
before in the work of Haldane \cite{haldane}, and Read and Sachdev
\cite{rs1,rs2}, and the subtle summation over the oscillating
Berry phases in Fig~\ref{fig6} is reviewed elsewhere
\cite{ssmott}. Here we recap the physical picture that emerges
from these studies. First, the Berry phases have been shown to
play an unimportant role for all configurations associated with
{\em smooth} changes of the N\'eel vector field. However they are
non-vanishing once we allow for singular `hedgehog' or monopole
configurations of the N\'eel field in space-time. These monopoles
correspond (in the quantum system) to tunnelling events which
involve a change in the skyrmion number, $\mathcal{Q}$, of the
spin configuration (see Eq. (\ref{defQ})). Of crucial importance
is the result that, due to the Berry phases, the skyrmion-number
changing operator (which corresponds to the monopole in the path
integral) transforms non-trivially under square lattice
translations and rotations.  As usual, the statistical weight in
the partition function must respect the symmetries of the problem.
This implies that (for $S$ not an even integer) the fugacity for
single monopoles oscillates with a particular pattern (with zero
average) on the lattice. Indeed, for $S=1/2$ the ``lowest''
skyrmion number-changing event with a non-oscillating fugacity
occurs for ``quadrupled'' processes in which $\Delta
\mathcal{Q}=\pm 4$.  The coincidence of this factor of $4$ and the
$4$-fold degeneracy of the VBS ground states is not accidental
(see below).

Now consider the description of the phases of the spin
Hamiltonian. In the N\'eel phase the monopole events are clearly
suppressed at low energies. However, in the paramagnet (the
`quantum disordered state of the N\'eel magnet) the space-time
configurations of the N\'eel field must be riddled with monopoles.
In other words, we may think of the paramagnet as having a
`condensation' of the skyrmion number changing operator. Now the
non-trivial transformation of this operator under lattice symmetry
operations (which is due to the Berry phases) leads to broken
lattice symmetry in the paramagnet - this may then be identified
as VBS order.

Note the dual role played by the monopole configurations. When
they proliferate N\'eel order cannot survive. At the same time
their proliferation induces broken translation symmetry. Actually
although we haven't explained it here, it can be shown that the
monopoles play a third, equally important, role. Their
proliferation leads to confinement of any $S=1/2$ quanta into
integer spin excitations. A similar confinement was illustrated in
Fig~\ref{fig3}b for the coupled dimer model, but here the `string
tension' of the confining potential is provided by the spontaneous
VBS order, rather than the modulation of bond strengths in the
Hamiltonian. The lowest lying confined $S=1$ excitation forms a
triplon particle which contributes to a $\chi_\varphi$ of the form
in Eq. (\ref{triplon}).

In our work\cite{shortpap}, we argued that a direct second order
transition is possible between these two phases, and that the monopoles
are asymptotically absent at long length and time scales at the critical
point. (More precisely, the oscillating single-monopole fugacities
average to zero in the continuum limit, and the remaining {\sl
  quadrupled} monopole fugacity renormalizes to zero at the critical
fixed point). We dubbed this a deconfined quantum critical point. The
absence of monopoles at the critical fixed point has the immediate
consequence that the skyrmion number $\mathcal{Q}$ is {\em conserved}.
This is an extra topological conservation law that is a special property
of the deconfined quantum critical point. It does not obtain away from
the critical point in either phase. It also does not hold at the
conventional O(3) fixed point that describes the transitions of
Section~\ref{lgw}. It thus provides a sharp distinction between the
deconfined critical point and the more conventional LGW theory ${\cal
  S}_{\varphi}$ in Eqn. \ref{sp}.

Note that the absence of monopoles at the deconfined fixed point
suggests that spin-$1/2$ spinons get ``liberated'' right at the
critical point (after all, it is the proliferation of the
monopoles in the paramagnet that confines the spinons). More
detailed considerations led to the proposal \cite{shortpap} that
the deconfined critical point is described by the following
critical theory of the fractionalized $S=1/2$ $z_a$ quanta coupled
to a non-compact U(1) gauge field $A_\mu$:
\begin{eqnarray}
&& \mathcal{Z}_{\rm deconfined} = \int \mathcal{D} z_a (r, \tau)
\mathcal{D} A_{\mu} (r, \tau) ~~~~~~~~\nonumber \\
&&~~~\times \exp \Biggl( - \int d^2 r d \tau \biggl[
|(\partial_\mu -
i A_{\mu}) z_a |^2 + s |z_a |^2  \nonumber \\
&~&~~~~~~~~~ + \frac{u}{2} (|z_a|^2)^2 + \frac{1}{2e^2}
(\epsilon_{\mu\nu\lambda}
\partial_\nu A_\lambda )^2 \biggl]\Biggl). \label{zc}
\end{eqnarray}
Here $z$ is a two-component spin-$1/2$ complex spinor. The
parameter $s$ is tuned by varying $g$, and its value must be
adjusted so that $\mathcal{Z}_{\rm deconfined}$ is at its own
critical point. The gauge field $A_{\mu}$ is closely related to
the compact field $\mathcal{A}_{\mu}$ defined in Fig~\ref{fig6},
but it becomes effectively non-compact in the continuum limit
appropriate to the critical point.

The N\'eel vector field is a composite of the spinon fields
\begin{equation}
\vec \varphi = z^{\dagger}\vec \sigma z \label{phiz}
\end{equation}
This is just the well-known $CP^1$ representation of the unit
vector $\vec \varphi$. However, it is absolutely crucial that the
gauge field $A_\mu$ be regarded as {\em non-compact}. Indeed, it
is the non-compactness that allows for the conservation of the
skyrmion number that obtains at the critical point. To see this,
note that we can also express \cite{rajaraman} the conserved
$\mathcal{Q}$ as
\begin{equation}
\mathcal{Q} = \frac{1}{2\pi} \int dxdy \left(\partial_x A_y -
\partial_y A_x \right),
\end{equation}
and it is evident that this is strictly conserved by
$\mathcal{Z}_{\rm deconfined}$ so long as the gauge field is
non-compact. In contrast for a compact gauge theory, instanton
events which change the gauge flux by $2\pi$ are allowed which
then kills the conservation of $\mathcal{Q}$. It should now be
clear that these instanton events precisely describe the
skyrmion-number changing monopoles.

How are we to reconcile the absence of monopoles at the critical
fixed point with their supposed proliferation in the paramagnetic
phase? The answer is that although (in the presence of appropriate
Berry phases) monopoles are irrelevant at the critical fixed point
of $\mathcal{Z}_{\rm deconfined}$, they are {\em relevant} at the
paramagnetic fixed point of this theory. Indeed the paramagnetic
phase of $\mathcal{Z}_{\rm deconfined}$ is aptly described as a
U(1) spin liquid \cite{hermele} with a gapless deconfined `photon'
field $A_\mu$. This is {\em unstable} to the inclusion of
monopoles (unlike the critical fixed point). The resulting flows
away from this U(1) spin liquid lead to the VBS phase with broken
translation symmetry, and confined spinons. The structure of the
renormalization group flows is shown in Fig.~\ref{fig9} (a similar
structure of flows was discussed earlier \cite{rs2,gs,csy} for
SU($N$) quantum antiferromagnets for large $N$). Unlike the usual
O($n$) fixed point of Eq.~(\ref{sp}), here the initial flow away
from the critical fixed point is not toward a stable paramagnet
but rather toward the unstable U(1) spin liquid state.
\setlength{\currentfigwidth}{\figwidth}
\addtolength{\currentfigwidth}{3in}
\begin{figure}[tb]
\centering
\includegraphics[width=\currentfigwidth]{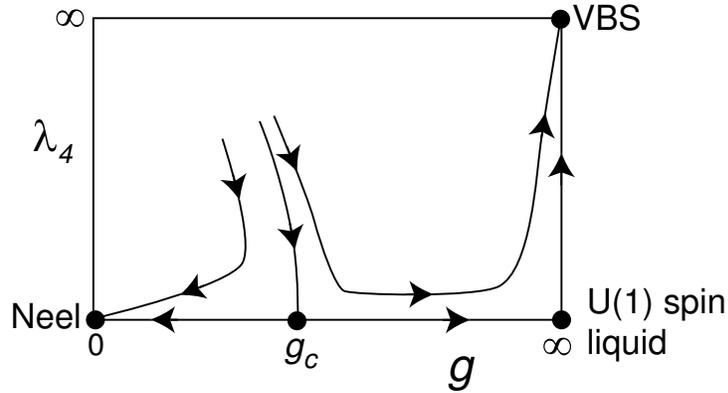}
\caption{Schematic renormalization group flows for the $S=1/2$
square lattice quantum antiferromagnet $H_s$ in
Eq.~(\protect\ref{hams}). The theory $\mathcal{Z}_{\rm
deconfined}$ in Eq.~(\ref{zc}) describes only the line
$\lambda_4=0$ (with $s \sim (g-g_c)$): it is therefore a theory
for the transition between the N\'{e}el state and a U(1) spin
liquid with a gapless `photon'. However, the lattice
antiferromagnet always has a non-zero bare value of the monopole
fugacity $\lambda_4$ (the monopoles are quadrupled by the Berry
phases, as reviewed elsewhere \cite{ssmott}). The $\lambda_4$
perturbation is irrelevant at the $g=g_c$ critical point of
$\mathcal{Z}_{\rm deconfined}$: this critical point therefore also
described the transition in the lattice antiferromagnet. However,
the $g\rightarrow \infty$ U(1) spin liquid fixed point is {\em
unstable\/} to $\lambda_4$, and the paramagnet is therefore a
gapped VBS state. In the earlier
discussion\protect\cite{rs2,gs,csy} of such flows for large $N$
SU($N$) quantum antiferromagnets, the monopoles were found to be
irrelevant at the critical point with or without Berry phases,
while for $N=2$, Berry phases are required to render the monopoles
irrelevant at criticality. It was this crucial distinction between
large and small $N$ which ultimately prevented a complete picture
from emerging from the early large $N$ studies \cite{rs2,gs,csy}.}
\label{fig9}
\end{figure}

Renormalization group flows with this structure have the general
consequence of having two distinct diverging length or time scales
(or equivalently two vanishing energy scales). Consider the
paramagnetic side close to the transition. First there is the spin
correlation length $\xi$ whose divergence is described by
$\mathcal{Z}_{\rm deconfined}$. At this scale there is a crossover
from the critical fixed point to the unstable paramagnetic U(1)
spin liquid fixed point which has the free photon. However the
instability of this spin liquid fixed point to VBS order and
confinement occurs at  a much larger scale $\xi_{VBS}$ which
diverges as a power of $\xi$.

The physical consequences of the existence of such a deconfined
critical point have been discussed in detail in our
work\cite{shortpap}. Here we simply make a few brief further
clarifying observations. One immediate consequence of the emergent
topological conservation law is that it fixes the scaling
dimension of the flux density (or N\'eel skyrmion density in terms
of the spin variables) operator $f_0= \left( \partial_x A_y -
\partial_y A_x \right)$. At criticality, this conservation law
implies $\langle f_0(R) f_0(0) \rangle \sim R^{-4}$ at long
distances. Furthermore, slightly away from the critical point this
conservation  of skyrmion number holds only up to the length scale
$\xi_{VBS}$ which diverges faster than the spin correlation
length.
The resulting structure of correlations close to the critical
point (on the VBS side) is shown in Fig~\ref{fig8}, and discussed
in detail in our work~\cite{shortpap}.
\setlength{\currentfigwidth}{\figwidth}
\addtolength{\currentfigwidth}{3in}
\begin{figure}[tb]
\centering
\includegraphics[width=\currentfigwidth]{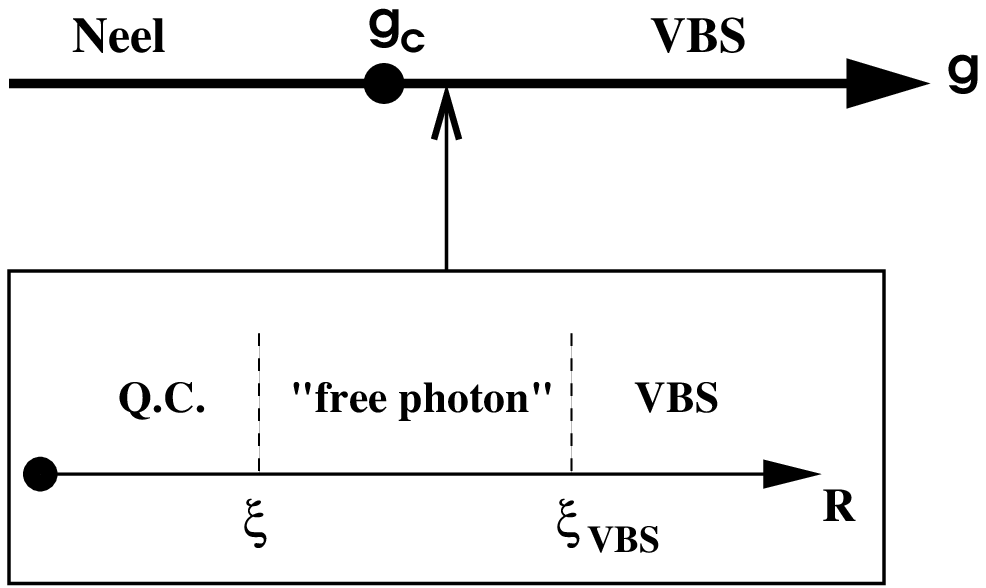}
\caption{Structure of correlations on approaching the deconfined
N\'eel-VBS quantum critical point from the VBS side. Two diverging
length scales, the correlation length $\xi$ and a longer length
scale $\xi_{VBS}$ are present.  As usual, at length scales $R$
shorter than $\xi$, quantum critical (Q.C.) correlations are
observed - {\em e.g.\/} spin-spin correlators are power law and
flux-flux correlators fall off as $\sim R^{-4}$. At intermediate
length scales, $\xi \ll R \ll \xi_{VBS}$, spin correlators are
exponentially decaying while flux-flux correlators take on the
free photon form $\sim R^{-3}$. At the longest length scales, only
VBS order is present.
 } \label{fig8}
\end{figure}
As usual, quantum critical correlations obtain for length scales
$R$ much smaller than the correlation length ($R \ll \xi$), where
for example spin-spin correlations show power law decay, while the
flux correlations have the $R^{-4}$ form described earlier. At
intermediate length scale $\xi \ll R \ll \xi_{VBS}$, the spin-spin
correlators fall off exponentially, but the flux density
correlators are power law but now decay as $R^{-3}$ which is
characteristic of flux correlations in the presence of free
photons. Finally at the longest length scales $\xi_{VBS} \ll R$,
VBS order is established and the photon is destroyed.

The continuum theory $\mathcal{Z}_{\rm deconfined}$ has a
strongly-coupled critical point, and the remarks made in
Section~\ref{lgw} for the critical field theory of
$\mathcal{S}_\varphi$ can be extended to the present situation.
The $z_a$ quasiparticles are not well defined at the critical
point, and characterized instead by their own anomalous dimension.
Indeed, the critical theory of $\mathcal{Z}_{\rm deconfined}$ may
be understood by the usual renormalized perturbative
analysis\cite{ma} but applied to a theory of nearly free,
fractionalized $z_a$ quanta. It is instructive to compute the
leading order prediction for the exponent $\eta$ in Eq.
(\ref{chicrit}) in such an approach. At tree level, the $z_a$
propagator is $1/p^2$ ($p$ is a spacetime 3-momentum); the
$\chi_\varphi$ susceptibility, by Eq. (\ref{phiz}), involves the
convolution of 2 such propagators, and so we obtain
\begin{equation}
\chi_\varphi (p) \sim \int \frac{d^3 p_1}{p_1^2 (p+p_1)^2} \sim
\frac{1}{|p|}
\end{equation}
Comparing with Eq. (\ref{chicrit}), this simple computation yields
a large anomalous dimension $\eta =1$. This illustrates our
claimed secondary characteristic of deconfined critical points:
the possibility of larger values for $\eta$.

In view of the crucial role played by the Berry phases in this
entire section, it is instructive to ask why they weren't a
serious issue in the coupled dimer model considered in
Section~\ref{sec:dimer}. The modulation of the coupling constants
in $H_d$ implies that  there is a natural pairing of the spins,
with each spin having a unique partner on its dimer.  This pairing
means that the Berry phase terms can be naturally grouped into
mutually cancelling terms. At sufficiently large scales the
cancelling Berry phases will renormalize to zero, and so their
effects can absorbed into effective values of the real terms in
the action; this was implicitly done in writing down the LGW field
theory in Eq.~(\ref{sp}). However, the intermediate length scale
over which the Berry phases have not cancelled out is also the
scale over which the constituent spinon-structure of the triplon
shown in Fig~\ref{fig3}b is present. This reasoning makes it clear
that the absence of explicit Berry phases in the LGW field theory
in Eq. (\ref{sp}) implies that this field theory contains no
signatures of spinon excitations, which contradicts the central
assertion of Bernevig {\em et al.} \cite{bgl}.

The natural pairing of the Berry phases into mutually cancelling
terms clearly does not extend to the model $H_s$ of interest in
this section. Each spin has a choice between four nearest neighbor
partners, and the choices between different spins are correlated
in a highly non-trivial manner.  So determining the appropriate
cancellations among the Berry phases is a much more delicate
matter\cite{shortpap,ssmott} and leads to the results described
above.

\section*{Acknowledgments}

We are grateful to R.~Shankar for valuable discussions and helpful
comments on the manuscript. This research was supported by the
National Science Foundation under grants DMR-0308945 (T.S.),
DMR-9985255 (L.B.), DMR-0098226 (S.S.)  and DMR-0210790,
PHY-9907949 (M.P.A.F.). We would also like to acknowledges funding
from the NEC Corporation (T.S.), the Packard Foundation (L.B.),
the Alfred P.  Sloan Foundation (T.S., L.B.), a John Simon
Guggenheim Foundation fellowship (S.S.), a Pappalardo Fellowship
(A.V.) and an award from The Research Corporation (T.S.).

\end{document}